\documentclass[journal=nalefd,manuscript=letter]{achemso}

\usepackage{float}
\usepackage{amsmath}
\usepackage{graphicx}
\usepackage{epsfig}
\usepackage{subfigure}
\usepackage{notes2bib}

\def\ai         {{\it ab initio}\;}
\def\eg         {{\it extended geometry}\;}
\def\cg         {{\it confined geometry}\;}

\def\BB         {{\bf B}}

\newcommand{\yambo} {{\normalfont\ttfamily yambo}}
\newcommand{\abinit} {{\normalfont\ttfamily abinit}}

\renewcommand{\[}{\left[}
\renewcommand{\]}{\right]}
\renewcommand{\(}{\left(}
\renewcommand{\)}{\right)}

\author{Davide Sangalli}
\email{davide.sangalli@fisica.unimi.it}
\affiliation[European Theoretical Spectroscopy Facility]{ETSF}
\affiliation[Dipartimento di Fisica, Universit\`a degli Studi di Milano,
via Celoria 16, I-20133 Milano, Italy]{Universit\`a di Milano, Italy}
\affiliation[Consorzio Nazionale Interuniversitario per le Scienze dei Materiali]
{CNISM}
\author{Andrea Marini}
\affiliation[European Theoretical Spectroscopy Facility]{ETSF}
\affiliation[Dipartimento di Fisica, Universit\`a di Roma ``Tor Vergata'',
via della Ricerca Scientifica 1, 00133 Roma, Italy]{Universit\`a di Roma ``Tor Vergata'', Italy}
\affiliation[IKERBASQUE, Basque Foundation for Science, E-48011 Bilbao, Spain]{IKERBASQUE Foundation, Spain}

\title[Anomalous Aharonov--Bohm gap oscillations in 
carbon nanotubes: a new perspective from an \ai approach]
{Anomalous Aharonov--Bohm gap oscillations in carbon nanotubes}
\begin{document}
\begin{abstract}
The gap oscillations 
caused by a magnetic flux penetrating a carbon nanotube represent
one of the most spectacular observation of the Aharonov--Bohm effect at the nano--scale.
Our understanding of this effect is, however, based on the assumption that the electrons
are strictly confined on the tube surface, on trajectories that are not modified by
curvature effects.
Using an \ai approach based on Density Functional Theory we
show that this assumption fails at the nano--scale inducing
important corrections to the physics of the Aharonov--Bohm effect.
Curvature effects and electronic density spilled out of the nanotube surface are
shown to break the periodicity of the gap oscillations.
We predict the key phenomenological features of this anomalous Aharonov--Bohm effect in
semi--conductive and metallic tubes and the existence of a large metallic phase in the
low flux regime of Multi-walled nanotubes, also suggesting possible experiments to
validate our results.
\end{abstract}

The Aharonov--Bohm(AB) effect\cite{Aharonov1959} is a purely quantum mechanical effect which does not have a
counterpart in classical mechanics. A magnetic field $\mathbf{B}$
confined in a closed region of space alter the kinematics
of charged classical particles only if they move inside
this region. Electronic dynamics,  instead, governed by
the Shr\"{o}dinger equation, is influenced even if the particles move on paths that enclose a
confined magnetic field, in a region where the 
the Lorentz force is strictly zero.
If these paths lie on the surface of a nanotube, 
electrons traveling around the cylinder are expected to manifest a shift of their phase.
The mathematical interpretation of this effect is connected with the definition of the vector 
potential, which, in the case of confined magnetic fields, cannot be nullified
everywhere.

This extraordinary effect, first predicted by Y.\,Aharonov and D.\,Bohm\cite{Aharonov1959}\,(AB)
in 1960, was interpreted as a proof of the reality of the electromagnetic potentials.
The idea that electrons could be affected by electromagnetic potentials
without being in contact with the fields was skeptically received by the scientific community.
At the same time the AB paper spawned a flourishing of experiments and extension of the original idea.
The first experiment aimed at proving (or disproving) the AB effect revealed a perfect
agreement with the theoretical predictions~\cite{Chambers1960}.
Nevertheless only some years later, in 1986, the experiment which can be considered as a
definitive proof of the correct interpretation of the AB effect was realized. 
Tonomura et al.\cite{Tonomura1986}, using superconducting niobium cladding,
were in fact able to completely exclude the possibility of stray fields as alternative
explanation of the predicted and observed AB oscillations.

Nowadays the AB effect can be used in a wide range of experiments, from the investigation
of the properties of mesoscopic normal conductors to the measure 
of the flux lines structure in superconductors. Growing interest is emerging in
the field of nanostructured materials. One of the most well-known case is
given by carbon nanotubes\,(CNTs) that, 
if immersed in a uniform magnetic field aligned with the tube axis, have been predicted
to show peculiar oscillations of the electronic gap~\cite{Charlier2007}. These oscillations are 
characterized by a period given by the magnetic flux quantum $h/e$ and 
are commonly interpreted as caused by the change in the wave 
functions of the electrons localized on the tube surface induced by the 
Aharonov-Bohm effect.

The first experiment carried on CNTs, in 1999, described the oscillations in the
electronic conductivity~\cite{Bachtold1999}, but with period of $h/2e$. 
This deviation from the predicted AB oscillation period has been explained in terms of
the weak localization effect~\cite{Abrahams1979} induced by defects and dislocation by
Al'tshuter, Aronov, Spivak~\cite{AAS1981}\,(AAS effect).
Only in 2004 a clear proof of the existence of the AB modulation of the electronic gap
with an $h/e$ period,
have been given by Coskun et coll.~\cite{Coskun2004} by measuring the conductance
oscillations in quantum dots. The dots were built using concentric
Multi--Walled CNTs of different radii, short enough to prevent the appearance
of weak localization. In the same year Zaric et al.~\cite{Zaric2004}
observed modulation in the optical gap of pure single walled CNTs 
with oscillations of $h/e$ period. 

Despite the enormous impact that the AB oscillations observed in CNTs had, their
explanation still remains grounded to simplified models,
like the Zone Folding approach (ZFA) or the Tight-Binding model (TBM)~\cite{Charlier2007}\,.
As CNTs are obtained by rolling a graphene sheet in different manners, the ZFA
assumes the electronic structure of the CNT to be well described by the
one of graphene.
The electronic states of the tube are defined to be the one of graphene allowed by
the boundary conditions imposed by the rolling procedure.
Magnetic field effects are then described as a shift of the allowed $k$ points
proportional to the magnetic flux~\cite{Charlier2007}.

Nevertheless both the ZFA and the TBM suffers from drastic limitations.
In the ZFA curvature effects are neglected as they decrease with increasing radii.
However it is well known that, at the nano--scale, the bending of the
electronic trajectories can induce effects of the same order of magnitude of the observed AB gap oscillations.
In the TBM CNTs can be described only introducing {\it ad--hoc} 
parameters, at the price of making the theory not quantitative and not predictive. 
In particular Multi--Wall\,(MW) CNTs, that are commonly produced experimentally, can be described, in the ZFA, only by 
neglecting the tube--tube interaction or, in the TBM, by introducing additional ad--hoc parameters~\cite{MWCNT-TB}.
More importantly, in both models the AB effect is introduced assuming that the electronic states are bi--dimensional,
strictly confined on the CNT surface, in contrast with the quantum nature of the electrons that, instead,
can induce tails in the wave--function that extend outside the CNT surface.

Therefore, the question we aim to answer is: do curvature effects and quantum spatial delocalization of the electronic
states induce quantitative modifications of the AB physics in CNTs\,?
In this work we perform an accurate and predictive study of the gap oscillations of single--wall and
multi--wall CNTs under the action of confined and extended magnetic fields.
By using a parameter--free approach based on Density Functional Theory we predict that, indeed, the gap oscillations 
are modified compared to the state--of--the--art understanding of the Aharonov--Bohm effect.
The key result is that in the standard experimental setup, 
when the CNT is fully immersed in an extended magnetic field, we predict 
non--periodic gap oscillations. 
We identify the non--periodic part of the oscillations as due to  the 
Lorentz force that, acting  on the electronic states spilled out of the CNT surface,
induces an additional gap correction.
In the case of the $(8,0)$ CNT we predict a non--continuous dependence of the gap, even in the low--flux regime,
induced by a metallic band that oscillates in the gap formed by $\pi/\pi^*$ states.
Another striking result we predict is the existence of a metallic phase induced by curvature effects in MWCNTs,
in the low-flux regime, that could be experimentally observed.
We show how the critical magnetic flux associated with this metallic phase increases when
the MWCNT is fully immersed in the magnetic field if compared to the case of a confined flux, thus providing
a clear finger--print of the geometry of the applied magnetic field.

We consider five CNTs: two metallic, two semi--conductive and one Multi--Walled. The ground--state of each
tube is computed at zero magnetic field within the local density approximation\,(LDA\cite{LDA}) using the plane--wave \abinit\
code~\cite{Calculation_details}.
The Magnetic field has been implemented self-consistently in the \yambo\, code~\cite{yambo}, which take as input
the LDA wave--functions and energies. The external magnetic field is added on top of the
DFT Hamiltonian $H_0$ by adding the correction $H_{magn}=\mathbf{A}^{ext}\cdot\mathbf{j}$.
Besides $H_{magn}$ also the correction to the non--local part of the
pseudo--potential~\cite{note_Vnl}
has been taken into account. The total Hamiltonian, then, is solved, self--consistently, in the DFT
basis~\cite{note_yambo_details}.

We simulate two different geometries of the applied magnetic field.
We consider a {\it confined geometry} where the magnetic field is confined inside the CNT and null outside, and
an {\it extended geometry} where the magnetic field is uniformly distributed.
The physical motivation of these different geometries traces back to the configuration of the applied magnetic field
commonly used in the experimental setups. Due to the difficulty of confining the magnetic field inside the 
CNT, a uniform field is commonly applied. 
It is worth to remind that in
the original experiment proposed by AB electrons move in a region which surrounds a confined magnetic
flux. Thus in the standard AB effect electrons travel in a space where {\bf B}$=${\bf 0}. In the experimental works, instead, the
confinement of the magnetic field is assumed to  be induced by the electronic trajectories confined on the tube surface.
As the concept of trajectory is, indeed, strictly valid only in the classical limit we deduce that the AB theory cannot be applied
straightforwardly to the standard experimental
setup~\cite{note_confined}.

\begin{figure}[H]
 \begin{center}
  \epsfig{figure=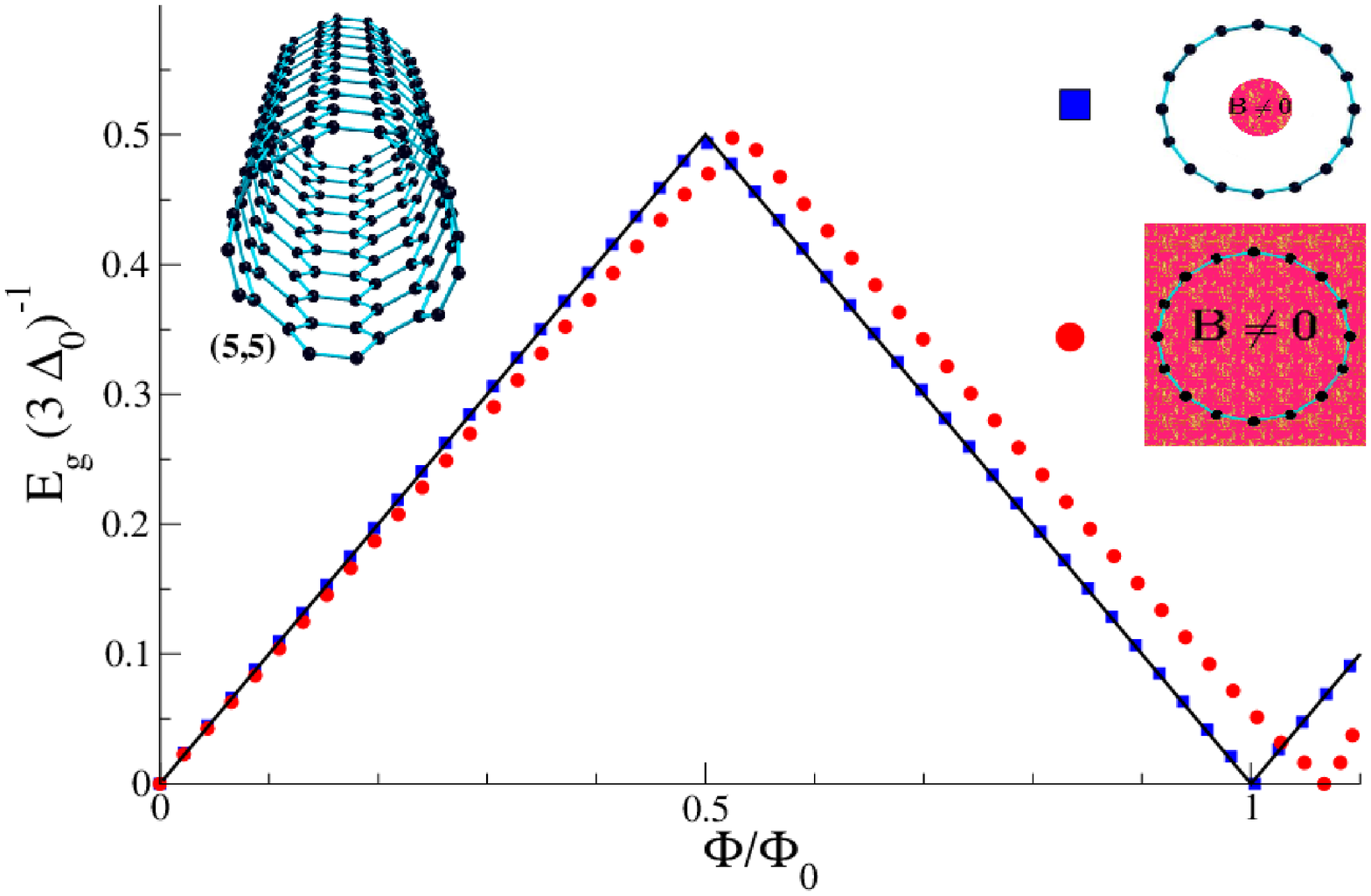,width=8cm}
  \epsfig{figure=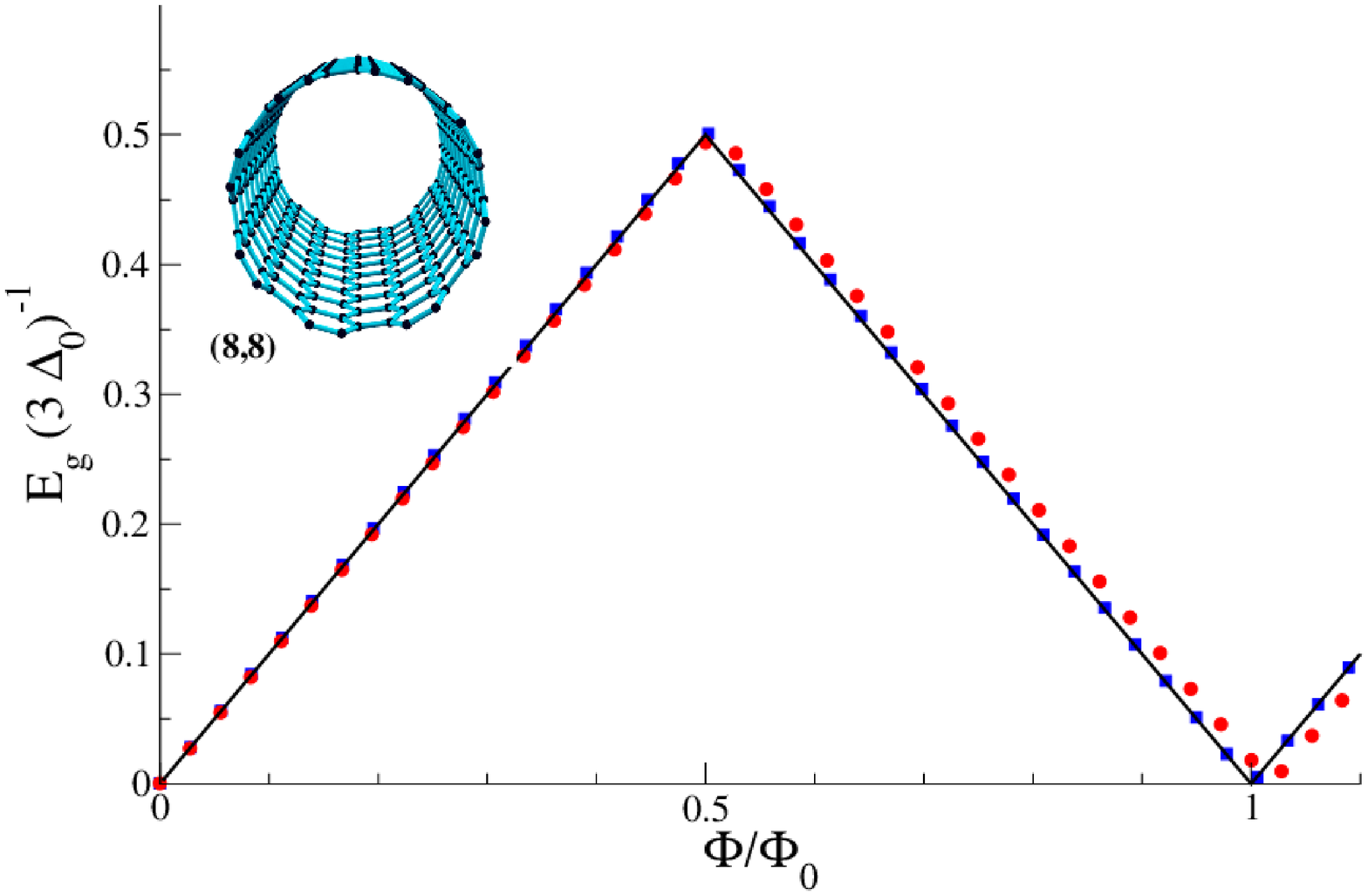,width=8cm}
 \caption{\footnotesize{Gap oscillations in the metallic (5,5) and (8,8) CNTs. Two different geometries
are considered. In the \eg the magnetic field is applied uniformly in all the space (red circles). In the \cg, instead,
the magnetic field is confined inside the CNT (blue boxes). We compare the \ai calculations with the ZFA results (black line).
We see that in the \eg, which represents the standard experimental setup, the Lorentz correction (see text) induces 
an overestimation of the elemental magnetic flux $\Phi_0$.}}
 \label{fig:CNT_gap_metallic}
 \end{center}
\end{figure}

First we consider two metallic CNTs: a (5,5) tube with radius 6.41\,Bohr and a (8,8) tube with
radius 10.22\,Bohr. In \ref{fig:CNT_gap_metallic}
we compare the gap dependence on the applied magnetic flux in the two  geometries with the result of the 
ZFA.
In the case of the smaller (5,5) tube we immediately see a first important deviation  of the \eg from the
\cg. The \eg,  which represents the standard experimental setup, 
overestimates by $\sim$7\% the elemental flux $\Phi_0$ which defines the periodicity of the 
gap oscillations.
This overestimation of the elemental flux alter the periodicity of gap. Indeed the correction induced 
in the \eg grows with the applied flux. From the dependence  of the elemental gap on the applied flux we may 
deduce that the overall electronic properties of the CNT are still periodic, although with a larger period, in
the applied flux. This is not true. As we will discuss later, in the \eg a non--periodic correction appears
that breaks the periodicity of the gap.

We want, first, to explain the different gap dependence obtained in the two geometries introducing, in
a formal manner, the Hamiltonian which governs the AB effect in the specific case of a CNT:
\begin{equation}
\label{eq:H}
H=-\frac{\hbar^2}{2m}\[ \frac{\partial^2}{\partial r^2}+\frac{1}{r}\frac{\partial}{\partial r}+
                            \frac{1}{r^2}\(i\frac{\partial}{\partial\varphi}-\frac{er}{\hbar}A\(r\)\)^2+
                            \frac{\partial^2}{\partial z^2}
                     \] + V(r,\varphi,z)
\text{.}
\end{equation}
Eq.\,\ref{eq:H} describes the electronic dynamics under the action of 
a static magnetic field, written in cylindrical coordinates centered on the axis of the CNT.
$A\(r\)$ is the vector potential which, in the symmetric gauge, describes a static magnetic
field along the $z$ direction and $V(r,\varphi,z)$ is the local DFT potential, which includes the 
ionic potential plus the Hartree and exchange--correlation terms.

\begin{figure}[H]
 \begin{center}
  \epsfig{figure=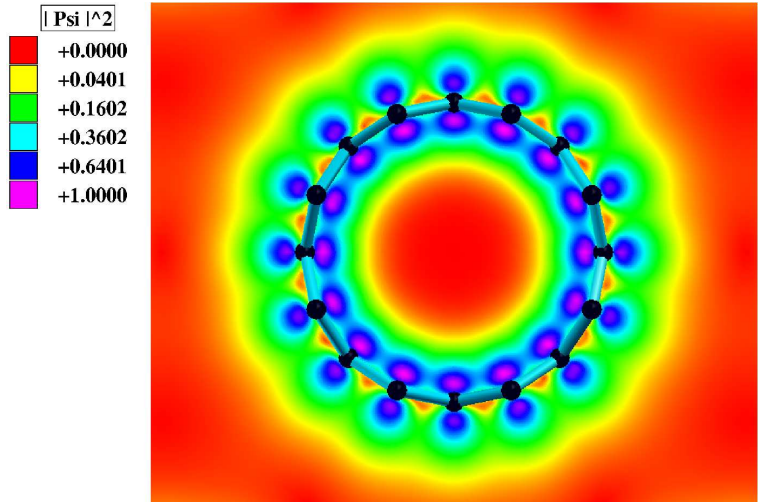,width=10cm}
  \caption{\footnotesize{Two dimensional plot of HOMO wave--function at the $\Gamma$ point for 
the $(8,0)$ CNT. The $\pi$ orbitals,
deformed by the  curvature of the CNT, have a larger amplitude in the outer part of the CNT surface.
As a consequence, in the \eg, electrons experience an effective flux which depends on their trajectory,
that is not localized on the CNT surface.
          }}
 \label{fig:wf_plot}
 \end{center}
\end{figure}

The only term of Eq.\,\ref{eq:H} which reflects the different geometry ({\it extended} or {\it confined}) is
$A\(r\)$. In the \eg
\begin{equation}\label{ext_pot}
A^{extended}\(r\)=\frac{1}{2}B_0 r,
\end{equation}
with $B_0=\left|\BB\right|$. In the \cg, instead, we have that
\begin{equation}\label{conf_pot}
A^{confined}\(r\)=\frac{1}{2}\frac{\Phi}{\pi r},
\end{equation}
with $\Phi=\pi B_0 R_{CNT}^2$ and $R_{CNT}$ the CNT radius. From Eqs.\,\ref{ext_pot},\ref{conf_pot} we see that
$A^{extended}\(R_{CNT}\)=A^{confined}\(R_{CNT}\)$, which implies that, if the electrons would exactly move on the
tube surface the \eg and the \cg would lead to the same gap oscillations. 
The different gap dependence observed in \ref{fig:CNT_gap_metallic}, is then due to the quantum nature of the electrons, that 
can spill out of the tube surface.
If we plug the two different expressions for $A\(r\)$ into Eq.~\ref{eq:H} we get two different
Hamiltonians, $H^{confined}$ and $H^{extended}$, whose difference is
\begin{equation}
\label{eq:H_corr}
\Delta H=H^{extended}-H^{confined}
        =\frac{e\hbar B_0}{2 m}\[
          \(1- \frac{R_{CNT}^2}{r^2}\) i\frac{\partial}{\partial\varphi}
          + \frac{eB_0r^2}{4\hbar} \(1- \frac{R_{CNT}^4}{r^4}\)           \]
\text{,}
\end{equation}
with $\Phi_0=h/e$. This term is zero when $r=R_{CNT}$, while near the 
tube surface behaves like $\sim \(1-\frac{R_{CNT}^2}{r^2}\)$. 
Now, as shown in \ref{fig:wf_plot}, the electrons are localized near, but not exactly {\it on} the tube surface.
Consequently $1-\langle R^2_{CNT}/r^2 \rangle\neq 0$ and, $\langle H^{extended}-H^{confined} \rangle \neq 0$.

We will refer to the correction defined by Eq. \ref{eq:H_corr} as Lorentz Correction (LC) as it 
introduces a magnetic term which depends on the electronic trajectory (through the term
$rA^{extended}_\phi$).
In the case of pure SWCNTs the LC appears, to first order, as an effective different radius 
of the electronic orbitals, as the correction would be zero defining the flux with respect
to the effective radius which satisfy the equation $1-\langle R^2_{eff}/r^2 \rangle= 0$ and so
an effective magnetic flux $\Phi'$.

From  \ref{fig:CNT_gap_metallic} we see that the ZFA matches the \ai simulation of the standard AB effect
corresponding to the {\it confined geometry} setup. This agreement is due to the fact that, in the ZFA
the LC is strictly zero as the electrons are assumed to move exactly on the graphite sheet, which represents
the CNT surface. Consequently in the ZFA  the electronic gap is function of the flux only.

Our results reveal that, if the flux is not confined inside the tube the electrons spilled out from the CNT surface
can alter the periodicity of the gap oscillations. More importantly the LC can {\em see} any deviation of the electron
orbital from the perfectly circular surface of the tube.
Consequently, even if the LC goes to zero in the case of SW--CNTs with increasing 
radius, impurities or defects can alter the electronic
trajectory creating deviations from a perfect circle of radius $R_{CNT}$. Even in large CNTs. 
We will see later a particularly large effect of the LC in the case of MWCNTs where bunches of electrons are
confined on tubes with different radii.

From \ref{fig:CNT_gap_metallic}, we may deduce that the LC increases the elemental flux $\Phi_0$ of the
gap oscillations, still maintaining the gap periodicity.
However this is not true. 
The gap periodicity is a distinctive feature of the standard AB effect. This is related to the gauge invariance
of the theory, that, in the \cg follows from the dependence of Eq.~\ref{conf_pot} on the magnetic flux only.
In the \eg, instead, Eq.~\ref{ext_pot} depends on the explicit position of the electron.
The LC, therefore, breaks the gauge--invariance of the theory
inducing a gap correction that is not periodic in the applied flux.

\begin{figure}[H]
 \begin{center}
 \subfigure{\includegraphics[width=7cm]{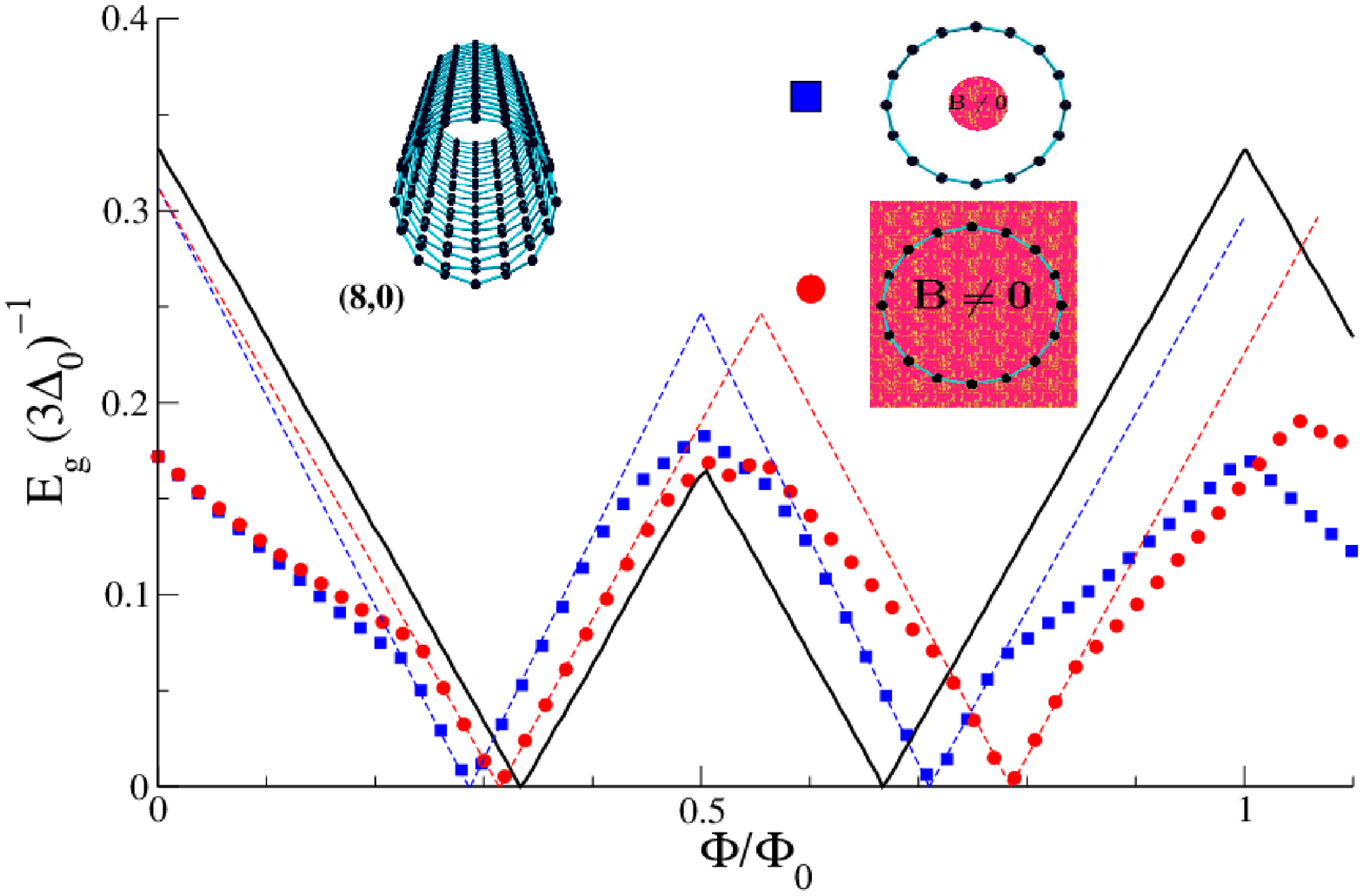} }
 \subfigure{\includegraphics[width=7cm]{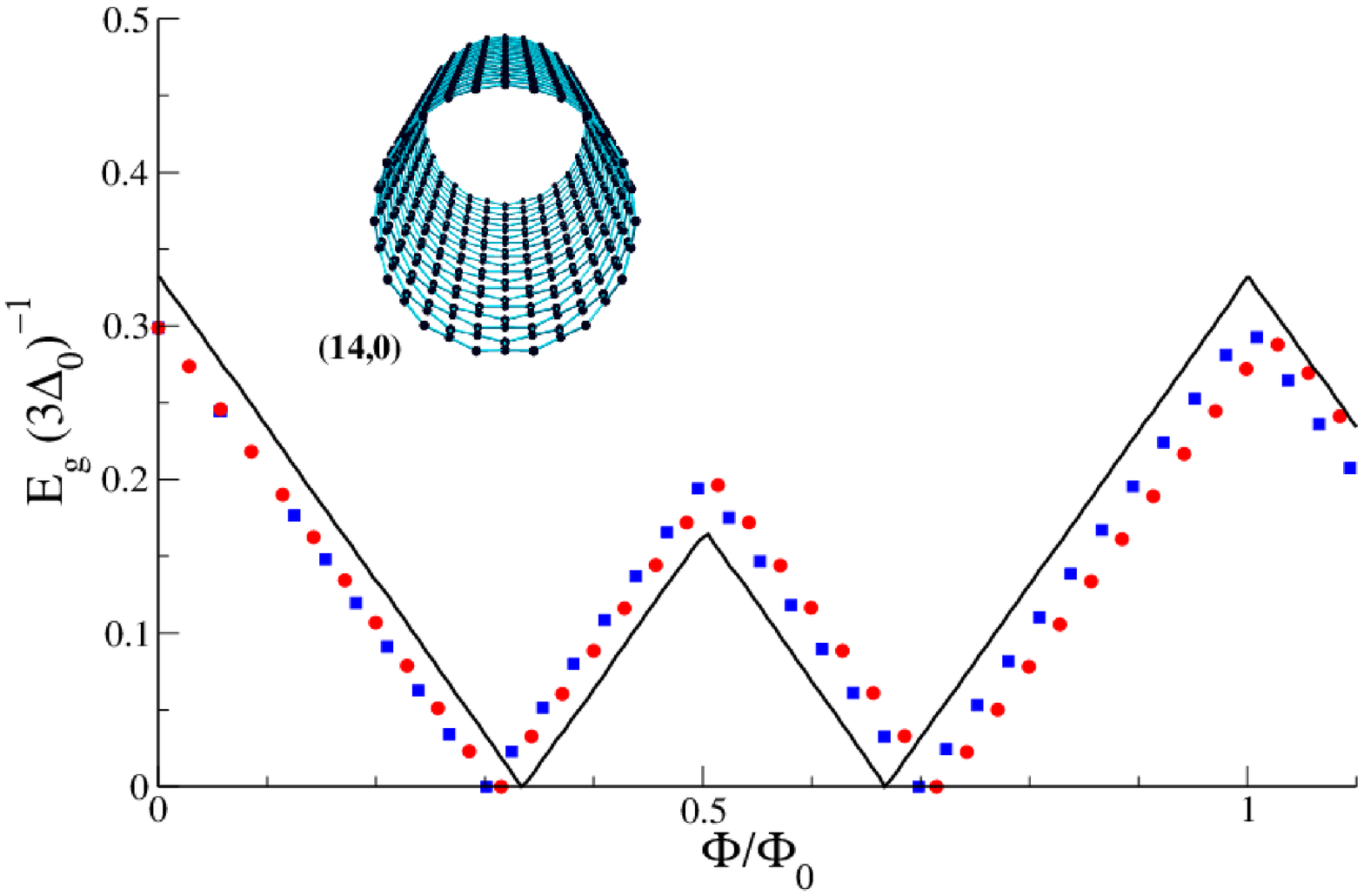} }
 \caption{\footnotesize{Gap oscillations in a semiconducting CNTs: (8,0) and (14,0). 
In the \eg the magnetic field is applied uniformly in all the space (red circles). In the \cg, instead,
the magnetic field is confined inside the CNT (blue boxes). We compare the \ai calculations with the ZFA results (black line).
In contrast to the
metallic case the curvature effects induce more evident differences between the {\it confined} and
the {\it extended} geometries.
In the (8,0)  tube the appearance of a metallic band in the gap deeply influences the gap behavior (see text). }}
 \label{fig:CNT_gap_semiconducting}
 \end{center}
\end{figure}

We now consider two semi-conducting CNTs: the (8,0) and the (14,0). 
The flux dependent electronic gap is shown in \ref{fig:CNT_gap_semiconducting}.
Similarly to the metallic case, LC makes the \eg to oscillate with a period
greater than $h/e$.
In contrast to the metallic case, the gap vanishes at two values of $\Phi$, which the
ZFA predicts to be at $\Phi_0/3$ and $2\Phi_0/3$, when the Dirac points becomes allowed
$k$ points~\cite{Charlier2007}. Noticeably both points are renormalized 
in the \ai simulation by curvature effects. 
It is well known, indeed, that, compared to graphene, curvature effects shift the Dirac points~\cite{Charlier2007}
$K$ at a position $|K|<2 \pi / 3a$, with $2 \pi / 3a$ the Dirac point position in graphene.
Accordingly a lower magnetic field is needed
to let the Dirac point belong to the set of the allowed $k$ points and
semiconducting CNTs becomes metallic at $\Phi<\Phi_0 / 3$.
Being the oscillations symmetric, the second metallic point is reached at $\Phi>2\Phi_0 /3$.

The deformation of the oscillations in the $(n,0)$ CNTs is smaller in 
bigger tubes. However it goes to zero slowly because both the shift and the magnetic period depend on the size
of the tube. For this reason the effect is still not negligible in the large $(14,0)$ tube as shown in
\ref{fig:CNT_gap_semiconducting}.

\begin{figure}[H]
 \centering
 \epsfig{figure=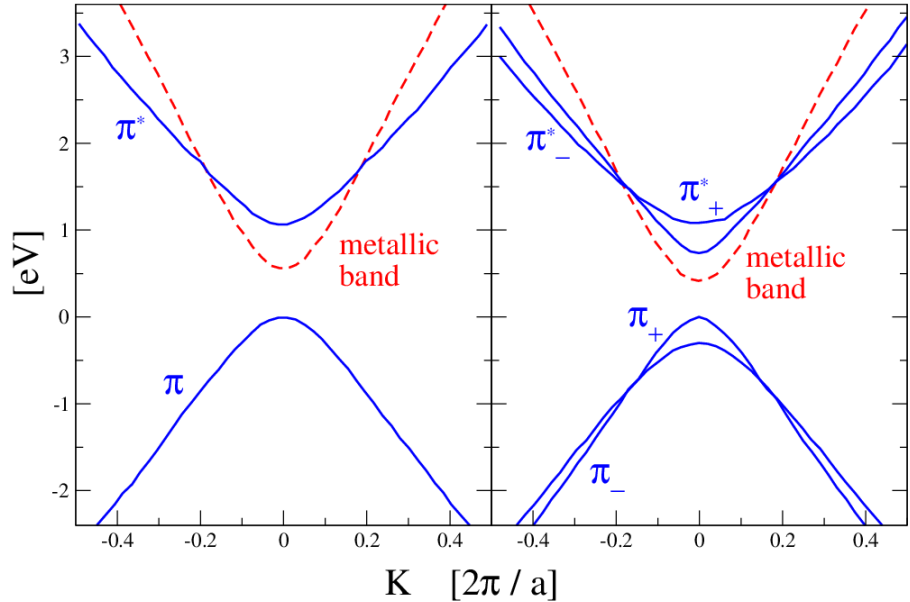,width=14cm}
 \caption{\footnotesize{Metallic--like (red dashed line) and $\pi/\pi^*$ (blue continuous line) bands of the $(8,0)$ CNT.
On the left panel $\Phi=0$, while on the right $\Phi>0$. 
At finite fluxes the metallic--like band is not split by the breaking of the time reversal symmetry.}}
 \label{fig:CNT_80_metallic_band}
\end{figure}

From \ref{fig:CNT_gap_semiconducting} we see that the $(8,0)$ gap
oscillations strongly deviate from the ZFA that does not reproduce, even qualitatively, the 
full \ai results. The reason for this large discrepancy traces back to the presence of a 
metallic--like band located near the Fermi surface.

This band is shown in \ref{fig:CNT_80_metallic_band} together with the $\pi / \pi^*$ bands close
to the Fermi level. When the $\mathbf{B}$ field is increased we see that, in contrast to the
$\pi / \pi^*$ bands, the metallic--like band does not shift, but moves inside the gap.
Consequently by changing the flux intensity the gap is defined by
transitions between the $\pi / \pi^*$ states or between the $\pi$ and the metallic--like band.
This explains the anomalous $\Phi$ dependence of $E_g$ shown in \ref{fig:CNT_gap_semiconducting}.

\begin{figure}[H]
 \begin{center}
 \epsfig{figure=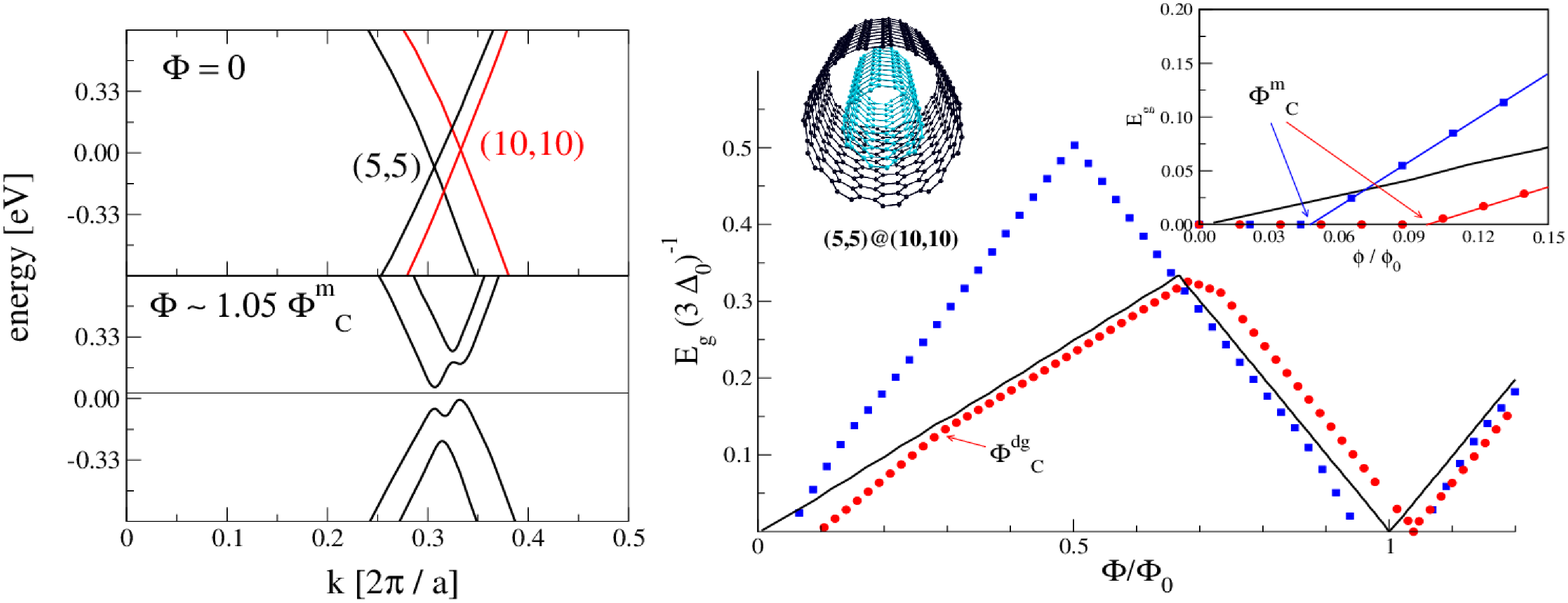,width=16cm}
 \caption{\footnotesize{Gap oscillations in a Multi--walled CNT: (5,5)@(10,10) (right frame). 
Zoom of the metallic intersection of the $\pi/\pi^*$ bands at zero and at low flux (left frames).
Two different geometries
are considered. In the \eg the magnetic field is applied uniformly in all the space (red circles). In the \cg, instead,
the magnetic field is confined inside the CNT (blue boxes). We compare the \ai calculations with the ZFA results (black line).
We immediately see that, in contrast to the ZFA, both geometries
are characterized by a metallic phase at low fluxes, highlighted in the inset.  The flux which
defines the end of the metallic phase is defined as $\Phi^m_c$. For $\Phi>\Phi^m_c$ the MW--CNT is an indirect 
semiconductor. For $\Phi=\Phi^{dg}_c$ the tube turns in a direct--gap semiconductor (see text).
}}
 \label{fig:MWCNT_gap}
 \end{center}
\end{figure}

Although SW--CNTs are routinely synthesized, MW--CNTs still constitute the majority of cases
used in the experiments. In \ref{fig:MWCNT_gap} we consider the case of a $(5,5)@(10,10)$ CNT
with radii 3.39 and 6.78 \AA. In this case the \cg is implemented considering 
the flux $\Phi=\pi B_0 R_{(10,10)}^2$. This flux is roughly the same experienced, in the \eg,
by the electrons of the $(10,10)$ CNT.

There is here a clear difference between the \cg and the \eg. The former describes 
a pure AB effect, where all the electrons feel the same magnetic flux, while in the latter
electrons can move on tubes with different radii and it is not possible any more to define a unique
magnetic flux. In this case the ZFA can be applied only by considering
two different Hamiltonians, one for the $(5,5)$ and one for the $(10,10)$ and by neglecting the
tube--tube interaction. 

The gap calculated in the \eg follows the ZFA prediction except in the very low field regime (and near 
the first inversion point, $\Phi\approx \Phi_0$). Moreover, 
from \ref{fig:MWCNT_gap}, we notice that, in the \cg, $\Phi_0$ is $\approx 4\%$ larger then in the ZFA. This
renormalization is larger then in the isolated $(10,10)$ SW--CNT that, according to our results,
is $< 1\%$. This follows from the fact that in the smaller $(8,8)$ tube the renormalization is $\approx 1\%$ and it is expected
to further decrease increasing the CNT radius. The reason for the enhancement of the LC in MW--CNTs can be understood 
by noticing that, in this case, the inner tube attracts the electrons of the outer tube thus 
increasing the fraction of electronic charge that is spilled out of the tube surface.  This 
phenomena decreases the effective radius felt by the electrons, thus increasing the 
effect of the LC.

One of the most severe problems that prevent a full investigation of the AB effect in small nano--structures is the
need of huge magnetic fields. For example the magnetic field corresponding to the elemental flux of 
the $(10,10)$ tube is $\approx 2900$~T. Nevertheless the present results suggest possible physical phenomena that
can be experimentally observed. 
We have seen, indeed, that in the simple case of a DW--CNT the existence of
an inner tube increases the LC effect by more than a factor 4. 
Therefore the effect of the LC, although 
decreasing with increasing tube radius, can be enhanced by altering mechanically the trajectories of the electrons. 
This is what commonly happens in the case of junctions, defects, dislocations or deformed CNTs. In this last case,
for example, an elliptical section of the tube would introduce an angle $\phi$ dependence in Eq.\ref{eq:H_corr}. This
would contribute to the breaking of the gauge--invariance, thus enhancing the LC.

Another phenomena, predicted by the present \ai calculations, that can be experimentally investigated, is the 
metallic regime for $\Phi\leq\Phi^m_c$ observed in MW--CNTs.
This regime can be described only by using a TBM with ad--hoc parameters included, {\it a posteriori}, to mimic the 
tube--tube interaction~\cite{MWCNT-TB+B}. The present scheme, instead, in a parameter--free manner,
predicts a metallic phase and, in addition, a transition from an indirect to a direct semi--conductive phase.
The metallic phase is a consequence of the different chemical potential felt by the electrons
moving on the $(5,5)$ and the $(10,10)$ surfaces. 
The \ai calculations shows that there is a $\approx 0.09 eV$ shift between the two
chemical potentials, as shown in \ref{fig:MWCNT_gap} (left frames), where the band
structure of the tube is shown at $\Phi=0$ (upper frame),
and at $\Phi\approx 1.1 \Phi^m_c$ (lower frame). 

At zero magnetic flux we see two pairs of crossing bands
which can be identified as the $\pi-\pi^*$ bands of the $(5,5)$ and the $(10,10)$ CNTs respectively. The two 
crossing points are not aligned in energy so that when a magnetic flux is applied two
small direct gap opens but the CNT remains metallic as long as the tip of the $\pi^*$ band of the 
$(5,5)$ is lower in energy than the one of the $\pi$ band of the $(10,10)$ CNT.
Only when $\Phi=\Phi^m_c$ a gap opens and the system turns in an indirect--gap
semiconductor, where the last occupied is a $\pi$ band of the outer (10,10) tube, while
the first unoccupied is a $\pi^*$ band of the inner (5,5) tube.

From the right frame of \ref{fig:MWCNT_gap} we also notice a sudden change in the
gap velocity~\cite{note_bohr_magneton},
defined as $v_g\(\Phi\)=\frac{1}{\mu_B}\frac{dE(B)}{d B}$, when $\Phi=\Phi^{dg}_c$.
At this flux the MWCNT turns in a direct--gap semiconductor, while for $\Phi<\Phi^{dg}_c$ the tube
is an indirect--gap semiconductor. This happens because the $\pi$ band of
the (10,10) tube moves faster than the $\pi$ band of the (5,5) tube.
At $\Phi=\Phi^{dg}_c$ the relative ordering of the two bands is inverted and the gap
is determined by the $(5,5)$ $\pi-\pi^*$ bands, thus recovering the prediction of the ZFA.

The critical field $\Phi^m_c$ represents, experimentally, an accessible flux regime.
Indeed from our \ai approach $B^m_c\approx 180$\,T for the $(5,5)@(10,10)$.
By using the dependence of the electronic properties of the CNTs on the radius\cite{note_critical}
we can extrapolate, for example, $B^m_c\approx 20$\,T for a $(10,10)@(15,15)$ MW--CNT.

We have discussed the metal--semiconductor transition induced by the magnetic field in 
the maximally symmetric geometry of an ideally isolated
(n,n)@(m,m) CNT. Nevertheless the effect on the electronic properties
of the tube--tube interaction and of a
different relative orientation of the two CNTs that compose the DW--CNT is
known from both ab--initio simulations~\cite{MWCNT-DFT,note_new_functionals}
and TB model results~\cite{MWCNT-TB,MWCNT-TB+B}.
Calculations performed by using the TB model on the 
(5,5)@(10,10) CNT~\cite{MWCNT-TB} show that the main effect is
the appearance of four pseudo--gaps.
Nevertheless, the tube remains metallic due
to the shift in the chemical potential, and the behavior of the electronic states perturbed
by the external magnetic flux
is not modified~\cite{MWCNT-TB+B}. Therefore the overall effect of the 
tube--tube interaction and of the
different relative angular orientation of the two CNTs would be a shortening
of the metallic phase. This corresponds to a reduction of the
critical flux even below the one we have predicted for the ideal DW--CNT



In conclusion, using an \ai approach, we have predicted that the gap 
of SW and MW--CNTs interacting with a magnetic field is characterized
by anomalous  Aharonov--Bohm oscillations. 
In the standard experimental setup, when the CNT is fully immersed in a uniform magnetic field, the gap oscillations
cannot be interpreted in terms of the standard Aharonov--Bohm  effect. Also in the case of a really confined
magnetic field the quantum nature of the electrons induce corrections to the pure Aharonov--Bohm  
oscillations.
Curvature effects and metallic bands have been shown to induce an anomalous dependence of the gap 
on the applied flux, not predicted by the state--of--the--art theory based on the Zone Folding approach.
The  quantum delocalization of the electronic wave--functions has been shown to induce
classical Lorentz Corrections that break the periodicity of the gap.
In MW--CNTs we have predicted the existence of metallic, indirect--gap and direct--gap semi--conductive phases
with the length of the metallic phase directly connected to the geometry of the applied
field.
We have also discussed how this metallic phase can be 
measured by using realistically weak magnetic fields, and how the Lorentz correction can appear, in the \eg, any time
a dislocation, a defect or a tube deformation alter the circular trajectory of the electrons.

The present results do contribute to improve the state--of--the--art understanding of the Aharonov--Bohm 
effect at the nano--scale. They clearly bind a quantitative description of the gap oscillations
induced by a magnetic flux to a careful inclusion of quantum effects on the electronic trajectories and to an
\ai, parameter--free  description of the nano--tube electronic states.

\acknowledgement

This work was supported by the EU through the FP6 Nanoquanta NoE
(NMP4-CT-2004-50019), the FP7 ETSF I3 e-Infrastructure (Grant Agreement
11956). One of the authors (AM) would like to acknowledge support from 
the HPC-Europa2 Transnational collaboration project. One of the authors (DS) 
would like to thank prof. Giovanni Onida and prof. Angel Rubio for supporting
this work and for useful discussions and suggestions.


\bibliography{NL_arxiv_28062011}

\begin{mcitethebibliography}{26}
\bibitem[Cal()]{Calculation_details}
 The ground state calculations were performed using a plane--wave basis and
  norm--conserving Troullier--Martins pseudopotentials. The kinetic energy
  cut--off used is $25$\,Ha. The Brillouin zone was sampled using k--points
  grids aligned with the tube axis. After carefull convergence tests we used
  $10$ k--points (in the irreduceble Brillouin Zone) for the $(n,0)$, $20$
  k--points for the $(n,n)$ and $30$ k--points for the $(5,5)@(10,10)$ CNT.
  Only for the $(5,5)@(10,10)$ CNT the elemental gap was calculated by using
  more refined k--point grids in the non self--consistent calculation.\relax
\mciteBstWouldAddEndPunctfalse
\mciteSetBstMidEndSepPunct{\mcitedefaultmidpunct}
{}{\mcitedefaultseppunct}\relax
\EndOfBibitem
\bibitem[not()]{note_Vnl}
 In particular we used Eq. 9 of Ref.~\cite{Mauri_nl_psp}.\relax
\mciteBstWouldAddEndPunctfalse
\mciteSetBstMidEndSepPunct{\mcitedefaultmidpunct}
{}{\mcitedefaultseppunct}\relax
\EndOfBibitem
\bibitem[not()]{note_yambo_details}
 \yambo\ diagonalizes self--consistently the Hamiltonian in the basis--set of
  the KS--wavefunctions at zero magnetic field. On average we need to include
  at least 50 (up to 100) conduction bands in order to obtain a converged gap
  up to $10^{-2}\ eV$. Moreover, to improve convergence \yambo takes advantage
  of gauge freedom in the choice of the vector potential.\relax
\mciteBstWouldAddEndPunctfalse
\mciteSetBstMidEndSepPunct{\mcitedefaultmidpunct}
{}{\mcitedefaultseppunct}\relax
\EndOfBibitem
\bibitem[not()]{note_bohr_magneton}
 The Bohr magneton $\mu_B\approx 5.79\ 10^{-5}$\, eV/T is introduced in order
  to make $v_g$ dimensionless.\relax
\mciteBstWouldAddEndPunctfalse
\mciteSetBstMidEndSepPunct{\mcitedefaultmidpunct}
{}{\mcitedefaultseppunct}\relax
\EndOfBibitem
\bibitem[not()]{note_new_functionals}
 However, only recently, new DFT functionals have been proposed to correctly
  describe the Van Der Waals component of the tube--tube
  interaction~\cite{MWCNT-DFT_functional}.\relax
\mciteBstWouldAddEndPunctfalse
\mciteSetBstMidEndSepPunct{\mcitedefaultmidpunct}
{}{\mcitedefaultseppunct}\relax
\EndOfBibitem
\end{mcitethebibliography}

\end{document}